\documentstyle[11pt,newpasp,twoside,epsf]{article}
\def\edcomment#1{\iffalse\marginpar{\raggedright\sl#1\/}\else\relax\fi}
\reversemarginpar

\begin{document}
\title{A physically consistent model for X-ray emission by Seyfert 2 galaxies demonstrated on NGC 1068}
 \author{E. Behar, A. Kinkhabwala, M. Sako, F. Paerels, S.M. Kahn}
\affil{Columbia University, 550 W 120th St., New York, NY 10027}
\author{A.C. Brinkman, J. Kaastra, R. van der Meer}
\affil{SRON, Sorbonnelaan 2, 3548 CA, Utrecht, The Netherlands}
\begin{abstract}
Preliminary analysis of the X-ray spectrum of NGC 1068 obtained by the RGS spectrometer on board {\it XMM-Newton} is presented. A physically consistent model is developed in order to quantitatively describe the reprocessing of the central AGN continuum source into the discrete X-ray emission observed in Seyfert 2 galaxies. All the important atomic processes are taken into account, including photoexcitation, which has been neglected in some previous models. The model fits the high resolution NGC 1068 data very well, which implies that the contribution of hot collisional gas to the X-ray spectrum of NGC 1068 is negligible.  
\end{abstract}
\section{Introduction}
Recent high-resolution X-ray observations of Seyfert 2 galaxies with the grating spectrometers on board {\it XMM-Newton} and {\it Chandra} reveal an emission spectrum rich in discrete features. This finding supports the unified AGN model, according to which the continuum nuclear source in Seyfert 2 galaxies is completely blocked from our line of sight and we can see only the reprocessed light. To date, discrete emission has been detected in Markarian 3 (Sako et al. 2000), NGC 4151 (Ogle et al. 2000), and the Circinus Galaxy (Sambruna et al. 2001). NGC 1068, which is the brightest Seyfert 2 and therefore best suited for detailed spectroscopic studies, has been observed by all X-ray grating spectrometers: LETGS (Brinkman et al. 2001), HETGS (Ogle et al. 2001), and RGS (Kinkhabwala et al. 2001 and the present work). The conspicuous radiative recombination continuum (RRC) features observed in all Seyfert 2 galaxies verify the photoionized nature and low-temperature ($kT_e \sim$ few eV) of the X-ray emitting plasma.

Spectra emitted by photoionized plasmas have been occasionally modeled assuming that the spectral lines and RRC's are solely produced by recombination and subsequent radiative cascades (e.g., Liedahl et al. 1990, Sako et al. 1999, Porquet \& Dubau 2001). When comparing these pure-recombination models to the Seyfert 2 X-ray spectra, significant residuals are found, which are mostly in lines with high oscillator strengths that are also among the strongest lines in hot collisional plasmas. Sako et al. (2000) were the first to suggest that these lines arise from photoexcitation (resonant scattering), while Ogle et al. (2000, 2001) interpret them as evidence for hot ($\sim$keV) plasma confining the much colder photoionized gas. Neither explanation has been tested with a detailed model yet. It is only natural that in a strong radiation field environment, the same flux that ionizes the atoms (bound-free transitions) will also excite them (bound-bound). In fact, emission by recombination and subsequent cascades {\it without} photoexcitation would be possible only if the plasma is in a transient phase of cooling and recombining. However, as a general rule in ionization balance, both of these processes contribute appreciably to the emitted spectrum. The relative importance of photoexcitation, recombination, and collisional excitation can be assessed only through careful modeling of the various line (and RRC) intensities, while correctly accounting for the atomic state kinetics. The model suggested in this paper is similar in spirit to the models that have been employed to interpret the spectra emitted by laboratory laser-produced plasma (Doron et al. 1998) and stellar coronae (Behar, Cottam, \& Kahn 2001). Only here, we add the transitions induced by the radiation field, namely photoionization and photoexcitation. We focus on the unique spectroscopic signatures of photoexcitation and point out their importance in the NGC 1068 spectrum obtained with the Reflection Grating Spectrometer (RGS) on board {\it XMM-Newton}. The full analysis of the NGC 1068 RGS observation will be presented elsewhere (Kinkhabwala et al. 2001).

\section{The Atomic-State Kinetic Model}
For each ion, we use a steady-state model, which includes all of the excited levels of the emitting ion in addition to the ground level of the next charge state. For the simple case of H-like ions, that would be the $nl$ configurations in addition to the bare nuclei state. The atomic processes included in the model are radiative decays, radiative recombination (RR), photoexcitation (PE), and photoionization (PI). Additional processes, such as collisional excitations, de-excitations, or dielectronic recombination processes can be easily incorporated, but their effect in typical Seyfert 2 plasmas is negligibly small. Self absorption of the reprocessed light can be included, but has been found not to be important for NGC 1068. The electron energy distribution is assumed to be Maxwellian, corresponding to an electron temperature $T_e$. The set of rate equations for the density $n_i^{+q}$ in cm$^{-3}$ of an ion with positive charge $q$ in a level $i$ can be written as:
\begin{eqnarray}
\frac{d}{dt} n_i^{+q} = \sum_{j>i}n_j^{+q}A_{ji} + n_e\sum_{k>i}n_k^{+(q+1)}\alpha_{ki}^{RR}(T_e) + \sum_{j<i}n_j^{+q}R_{ji}^{PE} \nonumber \\ 
 - n_i^{+q} \left( \sum_{j<i}A_{ij} + \sum_{j>i}R_{ij}^{PE} + \sum_{k>i}R_{ik}^{PI} \right) = 0
\end{eqnarray}

The electron density is denoted by $n_e$ and $A$ represents the Einstein coefficient for spontaneous radiative emission. $\alpha ^{RR}(T_e)$ is the rate coefficient for RR. The rates $R^{PE}$ for PE are calculated from the illuminating photon flux $F(E)$, which is a function of the photon energy $E$ and is expressed in photons sec$^{-1}$ cm$^{-2}$ keV$^{-1}$,
\begin{equation}
R_{ji}^{PE} = \int F(E)\sigma_{ji}^{PE}(E)dE
\end{equation}
The cross section for PE from level $j$ to level $i$ is:
\begin{equation}
\sigma_{ji}^{PE}(E) = \frac{\pi e^2}{m_ec}f_{ij}\phi(E)
\end{equation}
Here, $e$ and $m_e$ are the electron charge and mass, $c$ is the speed of light, $f_{ji}$ is the oscillator strength of the line, and $\phi (E)$ is the line profile. Analogously, the rates $R^{PI}$ for PI are calculated using the PI cross section and the same flux $F(E)$. $F(E)$ is taken to be a power law $[F_0(E)]$ absorbed along the ionization cone. The optical depth $\tau (E)$ is calculated taking both line (PE) and edge (PI) absorption into account. Since with the RGS, we measure the spectrum integrated over the entire ionization cone of NGC 1068, we use for Eq. (1) the effective flux :
\begin{equation}
F(E) = F_0(E)\frac{1-exp[-\tau(E)]}{\tau(E)}
\end{equation}
For spatially resolved spectra obtained with the {\it Chandra} gratings, a different flux needs to be used for each region. We will present a more detailed description of the model in Kinkhabwala et al. (2001).

Level-by-level calculations are carried out for each ion that appears in the spectrum. High lying levels with $n$-values up to 7 are included explicitly in the model. The contribution of higher levels, both via RR and via PE towards these levels, is included by extrapolation. The set of Eqs. (1) is solved by normalizing the sum of all level populations to unity. The power law normalization (or, alternatively, the electron density) is tuned to give the appropriate ionization balance between the two adjacent ions, which reproduces the observed RRC to line ratios. The line intensities in number of photons emitted per second and per cm$^3$ are finally given by $n_i^{+q}A_{ij}$. The elementary atomic quantities used in this work, namely level energies, radiative decay rates, and photoionization cross sections, were obtained by means of the multi-configuration, relativistic HULLAC (Hebrew University Lawrence Livermore Atomic Code) computer package developed by Bar-Shalom et al. (2001).
\section{Signature of Photoexcitation in the High Rydberg Series Lines}
	Looking at a Rydberg series of lines, for instance the Lyman series, the ratios of high-$n$ lines to that of, say, Ly$\alpha$ provide a probe for PE. At high ionic column density, beyond that in which the absorption of Ly$\alpha$ is saturated, the high-order lines can still be enhanced by PE resulting in the increase of their intensity ratio to Ly$\alpha$. This effect is illustrated in the left panel of Fig. 1 for C$^{+5}$. At the point where Ly$\alpha$ saturates ($\sim10^{16} cm^{-2}$), the ratios rise, eventually reaching values of up to a few times higher than the low column-density values. As the column density increases further, the other members of the series progressively saturate as well. The actual column densities for which all of these saturations occur depend on $\phi(E)$ (essentially the turbulent velocity, taken here to be 800 km/sec), but the intensity ratio trend seen in Fig. 1 is generic to the Seyfert 2 scenario, in which photons from a power law source are reprocessed by means of PE and PI. Next to the theoretical ratio curves in Fig. 1, still in the left panel, the ratios measured for NGC 1068 are shown (asterisks). These ratios are clearly in the high ionic column density regime at about 10$^{18}$ cm$^{-2}$. On the right hand side of the left panel in Fig. 1, the same ratios are indicated for collisional conditions where C$^{+5}$ forms ($kT_e$ = 100 eV). Clearly, these are lower than the ratios for photoionized conditions, even at their low column-density limit. Hence, an additional collisional component will only worsen the agreement of the model with the observed ratios.
\begin{figure}
\plottwo{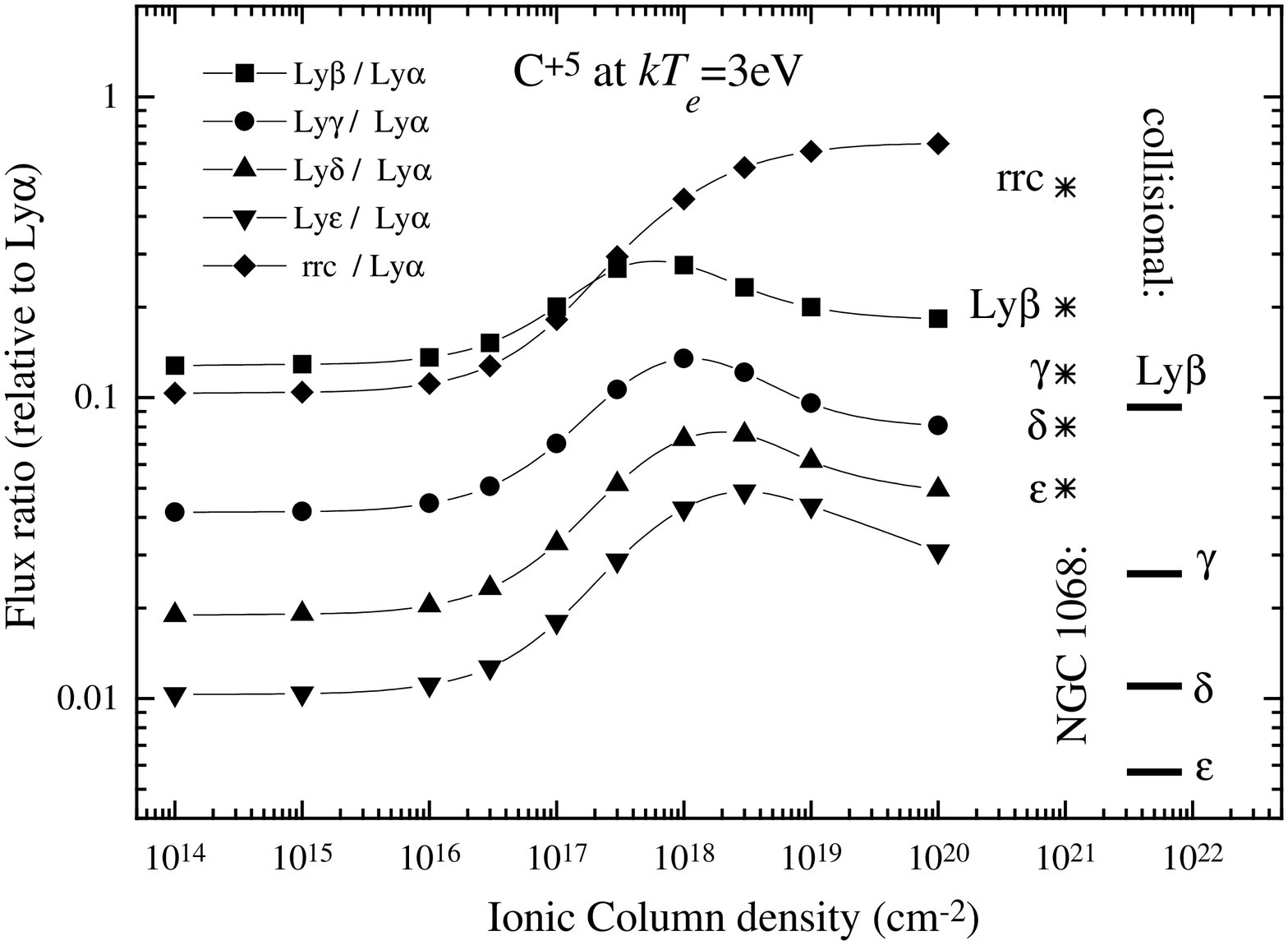}{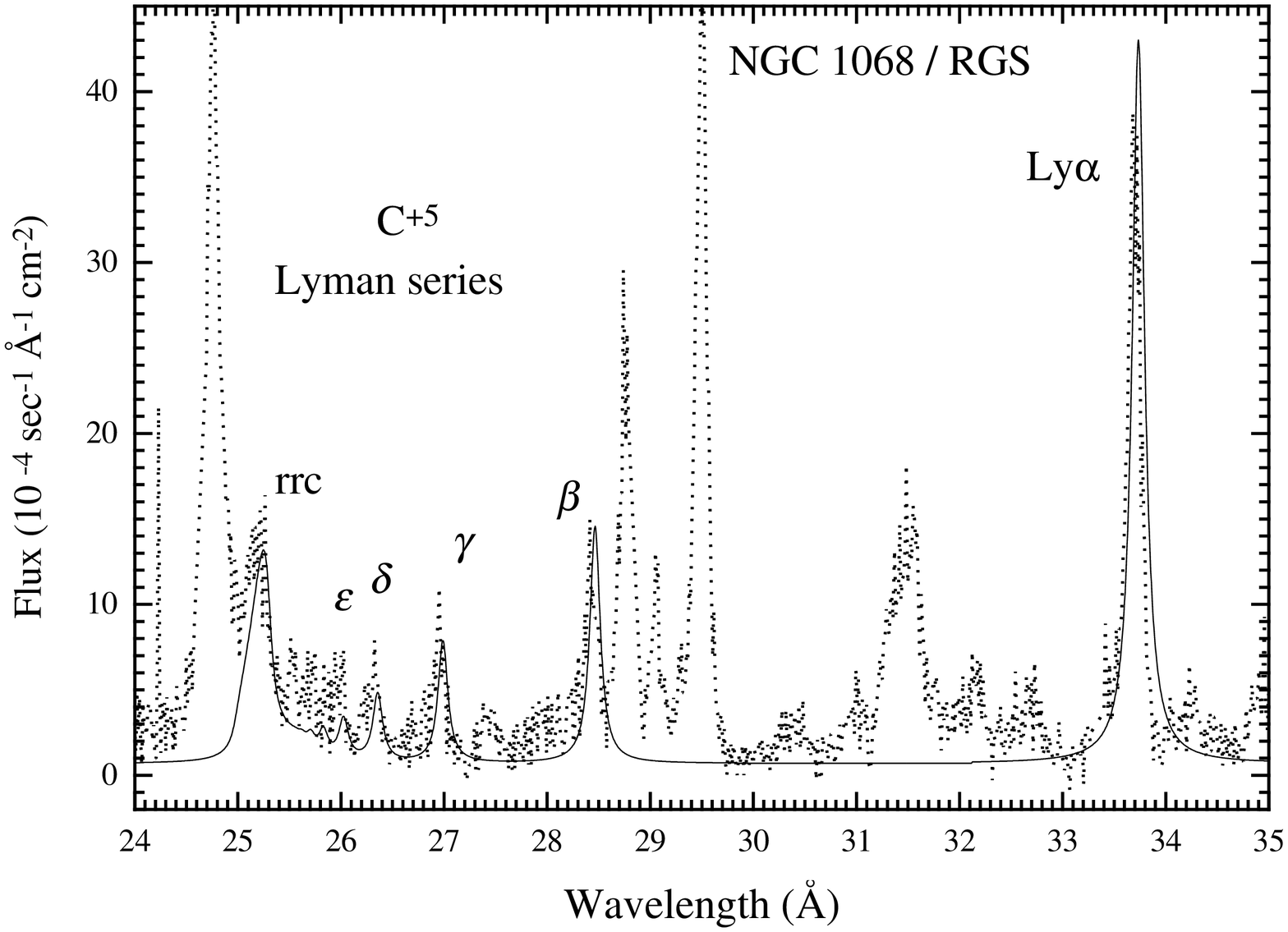}
\caption{{\it Left}- Calculated intensity ratios of the Lyman series lines and RRC (circles) of C$^{+5}$ to the Ly$\alpha$ intensity. The NGC 1068 values for these ratios, as obtained from the RGS spectrum, are also plotted (asterisks), as well as the values for collisional C$^{+5}$ gas at 100 eV (horizontal bars). {\it Right}- The NGC 1068 spectrum (dotted line) obtained by the RGS spectrometer on board {\it XMM-Newton} in the region of the C$^{+5}$ lines and RRC compared with the present model (solid line) at 10$^{18}$ cm$^{-2}$.}
\end{figure}

Indeed, the current, rather simple model for 10$^{18}$ cm$^{-2}$ fits the measured spectrum of NGC 1068 fairly well, as can be readily seen in the right panel of Fig. 1. The small remaining discrepancies could be due to the approximate line profiles used here. The complete analysis of the NGC 1068 spectrum, which includes all of the observed ions as well as the corrections for the line profiles and outflow velocity shifts will be presented in Kinkhabwala et al. (2001). Here, the slight velocity shifts actually come in handy and assist the comparison between the data and the model.
\section{He-like species}
The He-like species feature, in addition to the leading 1s - $n$p Rydberg series, the so-called forbidden ($f$) and intercombination ($i$) lines that together with the resonance ($r$) 1s - 2p line form the ($n$= 2 - 1; $J$ = 1 - 0) He-like triplet. For O$^{+6}$, these lines fall at 22.097, 21.804, and 21.602 \AA, respectively. Among these three, $r$ has the highest oscillator-strength by far and, thus, is most affected by PE. The intensity ratios of the various O$^{+6}$ lines to the $r$-line as a function of column density are plotted in the left panel of Fig. 2. An effect similar to that for the H-like species is seen for the high order lines ($\beta, \gamma, \delta$, etc.), where their intensity with respect to the $r$-line increases when the latter saturates, and reaching a plateau when they saturate as well. 
\begin{figure}
\plottwo{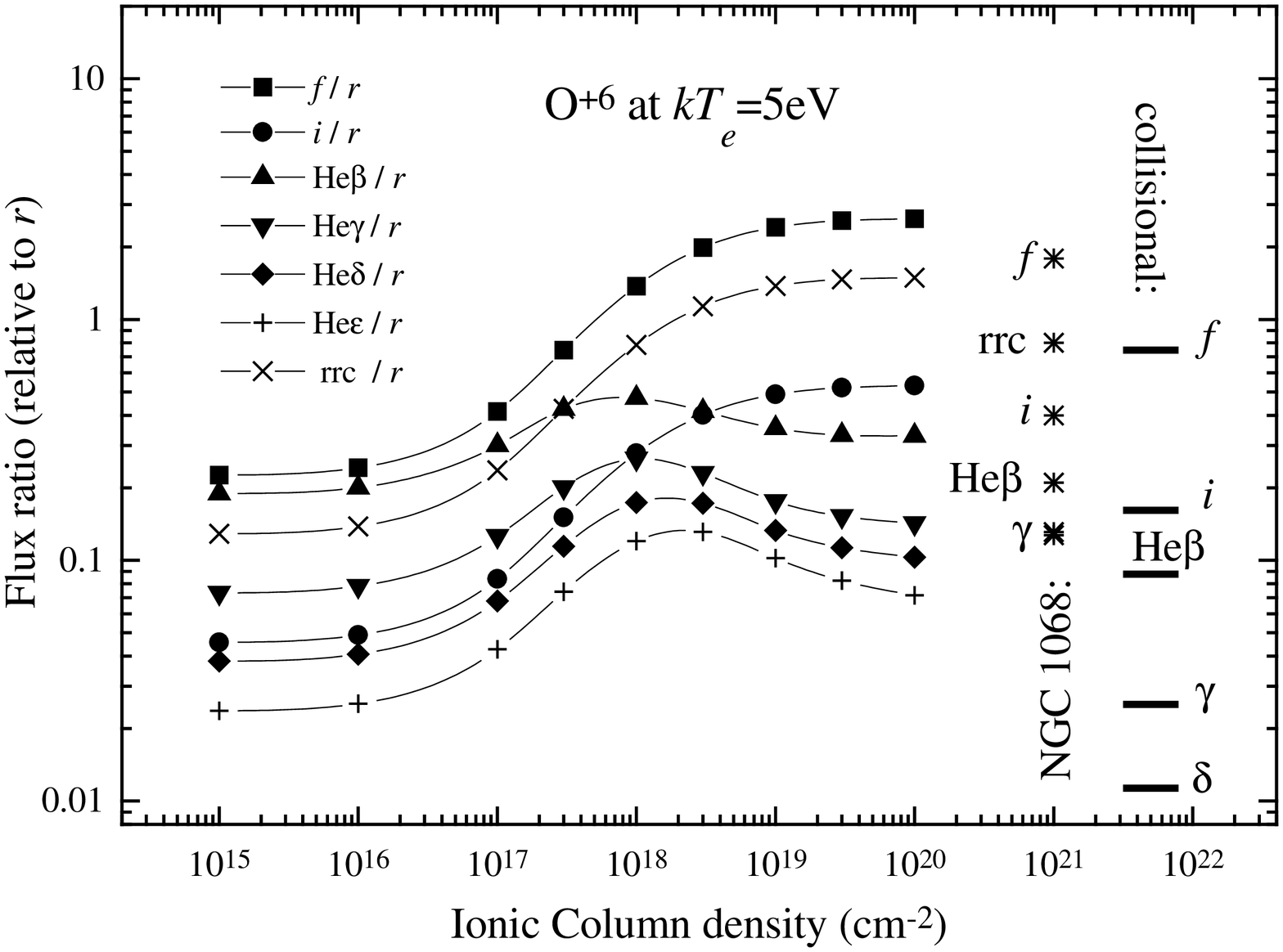}{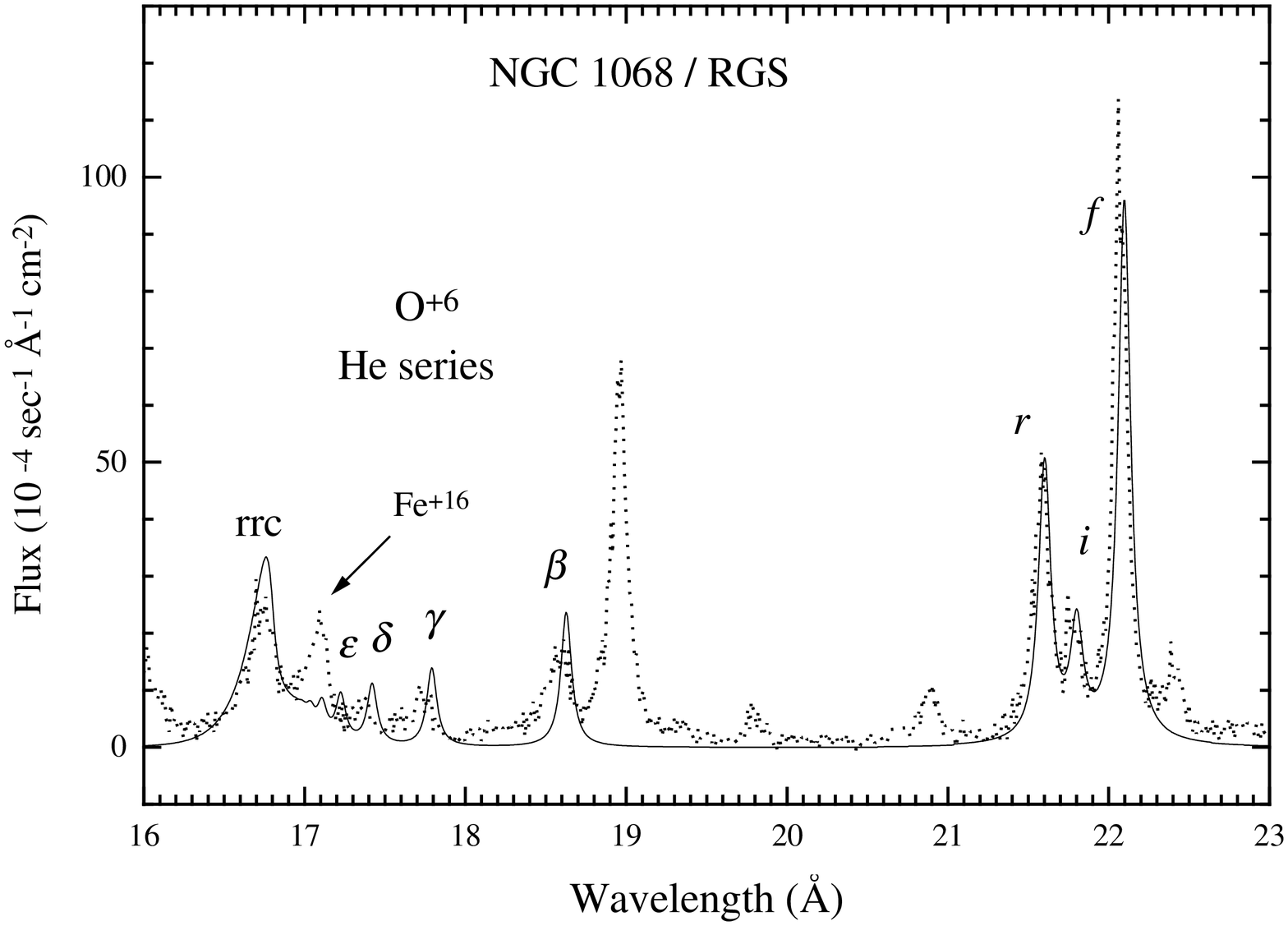}
\caption{{\it Left}- Calculated intensity ratios of the Lyman series lines and RRC (circles) of O$^{+6}$ to the Ly$\alpha$ intensity. The NGC 1068 values for these ratios, as obtained from the RGS spectrum, are also plotted (asterisks), as well as the values for collisional O$^{+6}$ gas at 100 eV (horizontal bars). {\it Right}- The NGC 1068 spectrum (dotted line) obtained by the RGS spectrometer on board {\it XMM-Newton} in the region of the O$^{+6}$ lines and RRC compared with the present model (solid line) at 10$^{18}$ cm$^{-2}$.}
\end{figure}
The line ratios for O$^{+6}$ measured from the NGC 1068 spectrum are also plotted in Fig. 2 (asterisks). The same as for C$^{+5}$, the measured O$^{+6}$ ratios are found to be higher than the collisional ratios, reducing the likelihood of hot collisional gas. The measured ratios of O$^{+6}$ imply a column density of $\sim 3\times 10^{18}$ cm$^{-2}$. The O$^{+6}$ model at this column density is plotted versus the data in the right panel of Fig. 2 and, evidently, the agreement is fairly good. 
\section{Conclusions}
We have shown that a physically consistent model for reprocessing of the nuclear continuum and subsequent emission from Seyfert 2 galaxies, which includes all of the important atomic processes and particularly PE, reproduces the high-resolution RGS X-ray spectrum of NGC 1068 very well. A more complete analysis of this spectrum is soon to appear in Kinkhabwala et al. (2001). Moreover, the traditional models, which did not include PE, clearly produce erroneous spectra and line ratios. This, in turn, could possibly lead to wrong astrophysical interpretation, such as the invoking of hot collisional gas, which to the best of our understanding is not evident from the X-rays emitted by Seyfert 2 galaxies.

\end{document}